%% file: main.tex
\documentclass[]{mosi}
\def\abstractlist{}
\usepackage{iftex}

\ifPDFTeX
\fi

\ifLuaTeX
  \pdfvariable objcompresslevel=0
\fi

\usepackage{helvet}

\usepackage{amsmath} 
\usepackage{natbib}
\usepackage{graphicx}
\usepackage{subcaption} 

\usepackage[toc,page,header]{appendix}
\usepackage[T1]{fontenc}    
\usepackage{hyperref}       
\usepackage{url}            
\usepackage{booktabs}       

\usepackage{amsfonts}       
\usepackage{nicefrac}       
\usepackage{microtype}      
\usepackage{wrapfig}

\usepackage{amssymb}        

\usepackage{titletoc}
\usepackage{minitoc}

\usepackage{array}
\usepackage{etoolbox}

\definecolor{lightblue}{RGB}{200, 230, 255}  
\definecolor{headerblue}{RGB}{150, 200, 255} 

\usepackage{pgfplots}
\usepackage{xcolor}
\usepackage{float}
\usepackage{comment}
\usepackage{multirow}
\usepackage{makecell}
\usepackage{siunitx}
\usepackage{tikz}
\usepackage{pgf-pie}
\usepackage[export]{adjustbox}

\usepackage{ragged2e}
\usepackage{tabularx}
\usepackage{caption}
\usepackage{enumitem}
\usepackage{pifont}
\usepackage[hang,flushmargin]{footmisc}

\usepackage{tcolorbox}
\tcbuselibrary{breakable}
\tcbuselibrary{skins}

\usepackage{listings}

\usepackage{threeparttable}
\usepackage{setspace}

\usepackage[scheme=plain]{ctex}

\definecolor{oursgray}{gray}{0.95}

\definecolor{MossCyan}{HTML}{82D9FF} 
\definecolor{MossBlue}{HTML}{82B1FF}

\definecolor{ForestGreen}{RGB}{34, 139, 34}
\definecolor{Red}{RGB}{255, 0, 0}

\definecolor{tickG}{HTML}{00C853}  

\definecolor{crossR}{HTML}{FF1744}

\definecolor{frenchblue}{rgb}{0.0, 0.45, 0.73}
\definecolor{babyblue}{rgb}{0.54, 0.81, 0.94}
\definecolor{classicrose}{rgb}{0.98, 0.8, 0.91}
\definecolor{beige}{rgb}{0.96, 0.96, 0.86}
\definecolor{forestgreen}{HTML}{2e7d43}

\definecolor{blue1}{HTML}{91BBE6}
\definecolor{blue2}{HTML}{3F90E0}
\definecolor{blue3}{HTML}{316FAD}

\definecolor{color1}{HTML}{FF9999}
\definecolor{color2}{HTML}{FF6666}
\definecolor{color3}{HTML}{FF3333}
\definecolor{color4}{HTML}{E60000}
\definecolor{color5}{HTML}{B30000}
\definecolor{color6}{HTML}{8CD98C}
\definecolor{color7}{HTML}{53c653}
\definecolor{color8}{HTML}{00B050}
\definecolor{color9}{HTML}{2d862d}
\definecolor{color10}{HTML}{206020}
\definecolor{color11}{HTML}{cca300}

\newtcolorbox{promptbox}[2][]{
    colback=white,
    coltext=black,
    arc=3mm,
    boxrule=0.5pt,
    colframe=black!60!white,
    title={#2},
    colbacktitle=black,
    coltitle=white,
    fonttitle=\bfseries,
    top=8pt,
    bottom=8pt,
    left=10pt,
    right=10pt,
    breakable,
    before upper={
        \linespread{1}\selectfont
        \setlength{\parskip}{1ex plus 0.2ex minus 0.2ex}
        \setlength{\parindent}{0pt}
    },
    #1
}

\title{OmniVAE: An Audio-Video VAE with Cross-Modal Alignment for Joint Generation}

\author{
\makebox[0.82\linewidth][s]{%
Jun Zhan$^{1,2,3,*,\ddagger}$\hfill
Chen Yang$^{1,2,3,*}$\hfill
Yitian Gong$^{1,3,*}$\hfill
Donghua Yu$^{1,3}$}
\\[0.35mm]
\makebox[0.88\linewidth][s]{%
Kuangwei Chen$^{1,3}$\hfill
Wenbo Zhang$^{1,3}$\hfill
Kexin Huang$^{1,3}$\hfill
Qi Luo$^{1,3}$\hfill
Zhe Xu$^{1,3}$}
\\[0.35mm]
\makebox[0.82\linewidth][s]{
Ying Zhu$^{1,3}$\hfill
Jin~Wang$^{1,3}$\hfill
Tengyue~Zhang$^{2,4}$\hfill
Qi Chen$^{2,4}$}
\\[0.35mm]
\makebox[0.88\linewidth][s]{%
Zhiyu Zhang$^{2,3}$\hfill
Cheng Chang$^{1,3}$\hfill
Songlin Wang$^{3}$\hfill
Junqi~Dai$^{1}$\hfill
Jiasheng~Ye$^{1}$}
\\[0.35mm]
\makebox[0.82\linewidth][s]{%
Xiaogui~Yang$^{3}$\hfill
Tianyi~Liang$^{2,3}$\hfill
Xiangyu~Peng$^{3}$\hfill
Zhaoye~Fei$^{1,2,3}$}
\\[0.35mm]
\makebox[0.88\linewidth][s]{%
Shimin Li$^{1,3}$\hfill
Qinyuan Cheng$^{1,3}$\hfill
Xie Chen$^{2,4}$\hfill
Xinchi Chen$^{1}$\hfill
Xipeng Qiu$^{1,2,3,\dagger}$}
\\[1.4mm]
{\normalfont\mdseries\fontsize{10}{12}\selectfont \texttt{%
$^{1}$Fudan University \quad
$^{2}$Shanghai Innovation Institute \quad
\\[0.6mm]
$^{3}$MOSI Intelligence \quad
$^{4}$Shanghai Jiao Tong University}}
\\[0.6mm]
{\normalfont\mdseries\fontsize{10}{12}\selectfont \texttt{%
$^*$Equal Contribution \quad $^{\ddagger}$Project Lead \quad $^{\dagger}$Corresponding Author}}
}

\checkdata[Homepage]{\url{https://openmoss.ai/OmniVAE.github.io}}
\checkdata[Code]{\url{https://github.com/OpenMOSS/OmniVAE}}
\checkdata[Model]{\url{https://huggingface.co/OpenMOSS-Team/OmniVAE}}

\input{chapters/abs}

\begin{document}
\maketitle
\begingroup
\renewcommand{\thefootnote}{\fnsymbol{footnote}}
\setcounter{footnote}{1}
\endgroup

\input{chapters/intro}

\begin{figure*}[t]
    \centering
    \includegraphics[width=\linewidth]{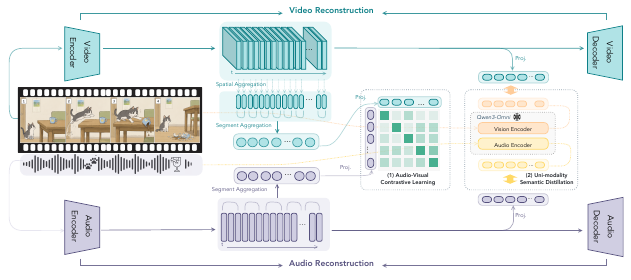}
    \caption{
    \textbf{Overview of OmniVAE.}
    OmniVAE keeps a separate video VAE (top) and audio VAE (bottom), each
    trained with its standard reconstruction objective (\textit{Video /
    Audio Reconstruction}). Beyond reconstruction, we introduce two
    training-only objectives on top of the latents:
    \textbf{(1) Audio-Video Contrastive Learning} aligns the two
    modalities in a shared space at fine temporal granularity---a spatial
    aggregator collapses each video latent frame into a token sequence,
    a segment aggregator pools video and audio latents into matched
    per-segment features, and a bidirectional InfoNCE loss is applied on
    the resulting similarity matrix.
    \textbf{(2) Uni-modality Semantic Distillation} supervises each
    branch with features from the frozen Qwen3-Omni~\citep{qwen3omni} vision and audio
    encoders, improving the learnability of each latent space on its own.
    The contrastive head and distillation projectors are used only during
    training; at inference, the two VAEs run independently with no added
    cost.
    }
    \label{fig:arch}
\end{figure*}

\input{chapters/related_works}

\input{chapters/method}

\input{chapters/exp}

\input{chapters/conclusions}

\bibliographystyle{unsrtnat}
\bibliography{main}


\end{document}

%% file: chapters/abs.tex
\abstract{
Recent generative models are moving beyond silent video or standalone audio synthesis toward the joint generation of synchronized audio and video. Despite this progress, jointly generating audio and video with fine-grained cross-modal correspondence remains challenging due to their fundamental structural differences. Most existing methods use audio and video VAEs trained separately. As a result, the two latent spaces lack cross-modal alignment, leaving the downstream generative model to learn cross-modal synchronization from scratch.
We present \textbf{OmniVAE}, a jointly trained audio-video VAE that learns fine-grained semantic alignment between audio and video latent representations. Beyond reconstruction, OmniVAE uses a segment-level audio-video contrastive objective to capture temporal-semantic correspondence and align the two latent spaces. In parallel, it distills features from pretrained modality-specific semantic encoders into each modality, improving the downstream learnability of both latent spaces.
Extensive experiments show that both objectives consistently improve the learnability of the latent spaces, translating into higher generation quality and more accurate cross-modal synchronization in downstream text-to-audio-video generation. These findings underscore the importance of learning unified representations as a foundation for omni-modal modeling.
}

%% file: chapters/intro.tex
\section{Introduction}

Unified audio-video generation aims to produce a video and its temporally synchronized, semantically coherent soundtrack jointly from a single text prompt, rather than generating the two modalities separately and combining them afterward. However, existing systems typically rely on independently trained, modality-specific VAEs~\citep{kingma2014autoencoding}. A tokenizer defines the representation space in which generation is learned, yet reconstruction alone does not ensure that this space is semantically structured or cross-modally aligned. The downstream generator must therefore learn both the data distribution and the correspondence between audio and video, making fine-grained synchronization particularly challenging.

Recent studies point to two complementary ways of improving this representation space. A growing body of work enriches modality-specific tokenizers with semantic structure: VA-VAE~\citep{yao2025vavae} and RAE~\citep{zheng2025rae} improve visual latent representations using pretrained vision encoders, while SpeechTokenizer~\citep{zhang2024speechtokenizer} and MOSS-Audio-Tokenizer~\citep{mossaudiotokenizer2026} pursue semantically meaningful audio representations. Meanwhile, video-to-audio methods such as MMAudio~\citep{cheng2025mmaudio} and HunyuanVideo-Foley~\citep{shan2025hunyuanvideofoley} demonstrate that explicit synchronization features~\citep{iashin2024synchformer} substantially improve temporal precision, highlighting the value of cross-modal alignment.

Motivated by these advances, we aim to make audio-video latent representations more learnable for downstream generative models by embedding modality-specific semantics and cross-modal correspondence directly into the VAE latent spaces. We therefore present \textbf{OmniVAE}, which learns aligned modality-specific latent spaces for audio and video while retaining the reconstruction capability of each modality. Its training is guided by two complementary objectives beyond reconstruction. A \emph{modality-specific semantic distillation} objective injects semantic structure into each latent space using frozen pretrained encoders, while a \emph{segment-level bidirectional InfoNCE} objective~\citep{oord2018cpc} aligns audio and video latents at fine temporal granularity. Both objectives are training-only and incur no additional inference cost. Experiments show that the resulting representations are more learnable both within individual modalities and jointly across audio and video, leading to higher generation quality and more accurate audio-video synchronization in downstream text-to-audio-video generation.

Our contributions are threefold. \textbf{(1)} We present \textbf{OmniVAE}, a jointly trained audio-video VAE that learns aligned modality-specific latent spaces with fine-grained semantic and temporal correspondence across modalities. To the best of our knowledge, OmniVAE is the first audio-video VAE designed explicitly for cross-modal latent alignment. \textbf{(2)} We introduce two complementary training objectives: modality-specific semantic distillation improves the learnability of each latent space, while segment-level audio-video contrastive learning aligns the two modalities at fine temporal granularity. Both objectives are training-only and incur no additional inference cost. \textbf{(3)} Through latent-space probing and downstream text-to-audio-video generation, we demonstrate that these objectives improve complementary aspects of the representation. Their combination consistently delivers the strongest overall generation quality and audio-video synchronization.

%% file: chapters/related_works.tex
\section{Related Work}


\subsection{Multimodal Unified Representation}

We use \emph{multimodal unified representation} to refer to a representation scheme in which different modalities are organized within a common semantic coordinate system, allowing semantically corresponding content to be directly related across modalities. Such unification does not require a shared encoder or identical latent distributions; it can be achieved through either a shared representation space or explicitly aligned modality-specific spaces.

Prior work explores such unification at different levels. Global contrastive methods organize heterogeneous modalities within a shared semantic space~\citep{radford2021clip,girdhar2023imagebind,zhu2024languagebind}, while discrete representation learning seeks modality-agnostic codes that support cross-modal transfer~\citep{liu2022crossmodal,xia2023cmg,huang2025enhancing}. These approaches capture high-level semantic correspondence but provide limited modeling of fine-grained temporal relationships. In contrast, SyncNet~\citep{chung2017syncnet} and Synchformer~\citep{iashin2024synchformer} explicitly learn temporal correspondence between audio and video for synchronization.

However, these representations are primarily designed for discriminative tasks and are generally not optimized to reconstruct the original signals. They therefore cannot directly serve as latent spaces for generative modeling. OmniVAE bridges this gap by learning decodable audio and video representations that preserve modality-specific reconstruction while establishing fine-grained cross-modal alignment.

\subsection{Audio and Video Tokenizers}

Modern latent generative models rely on tokenizers that compress high-dimensional raw signals into compact latents on which generation is performed. On the video side, continuous VAEs under the latent-diffusion paradigm (e.g., Wan-VAE~\citep{wan2025} and the Movie Gen VAE~\citep{polyak2024moviegen}) compress pixels into a continuous latent through spatio-temporal downsampling; on the audio side, neural codecs and continuous VAEs (e.g., DAC~\citep{kumar2023dac}, the Stable Audio VAE~\citep{evans2024stableaudio}, and Qwen-Audio-VAE~\citep{jiang2026qwenaudiovae}) play an analogous role. Qwen-Audio-VAE further scales this reconstruction-oriented paradigm to low-bitrate, high-throughput general-audio encoding. Such tokenizers are traditionally trained with reconstruction (plus adversarial) objectives alone, so their latent spaces lack explicit semantic structure. A recent line of work instead injects semantics \emph{into the tokenizer itself}. On the visual side, recent work enriches generative tokenizers with semantic representations through feature alignment (VA-VAE~\citep{yao2025vavae} and REPA-E~\citep{leng2025repae}), direct reuse of pretrained foundation encoders (RAE~\citep{zheng2025rae}), or joint reconstruction and semantic pretraining (UniTok~\citep{ma2025unitok} and VTP~\citep{yao2025vtp}). Collectively, these studies show that semantic structure, rather than reconstruction fidelity alone, is critical to the learnability of visual latent spaces. On the audio side, SpeechTokenizer~\citep{zhang2024speechtokenizer} and MOSS-Audio-Tokenizer~\citep{mossaudiotokenizer2026} similarly integrate semantic learning into audio tokenization, seeking representations that preserve acoustic fidelity while better supporting downstream generative modeling. Nevertheless, these semantic tokenizers remain modality-specific: they improve the representation of each modality independently but do not establish correspondence across modalities. To the best of our knowledge, OmniVAE is the first semantic tokenizer that both distills modality-specific semantics into audio and video latents and explicitly aligns the two latent spaces at the segment level.

\subsection{Audio-Video Generation}

Audio-video generation aims to jointly synthesize video and audio that are semantically consistent and temporally synchronized. MM-Diffusion~\citep{ruan2023mmdiffusion} first proposed a coupled dual-U-Net diffusion framework with random-shift attention to jointly denoise video frames and waveforms, and AV-DiT~\citep{wang2024avdit} extends this idea to Diffusion Transformer (DiT) backbones. Industrial systems such as Sora~2~\citep{sora2}, Veo~3~\citep{veo3}, and Seedance~2.0~\citep{seedance2} now natively generate video with synchronized audio at scale. On the open-source side, JavisDiT~\citep{liu2025javisdit} introduces a hierarchical spatio-temporal synchronization prior inside a joint DiT, while Ovi~\citep{low2025ovi}, LTX-2~\citep{hacohen2026ltx2}, and MOVA~\citep{mova} all couple a video backbone and an audio backbone in twin-tower (dual-stream) designs through bidirectional cross-modal attention for synchronized audio-video generation. A closely related line injects synchronization features as \emph{external conditioning} into the generator: MMAudio~\citep{cheng2025mmaudio} feeds Synchformer-derived~\citep{iashin2024synchformer} frame-level visual features into a video-to-audio flow-matching transformer through AdaLN modulation and reports roughly a $50\%$ relative improvement in synchronization, and HunyuanVideo-Foley~\citep{shan2025hunyuanvideofoley} likewise relies on Synchformer features for gated temporal modulation---both underscoring the importance of explicit cross-modal alignment signals. In all of these systems, the video and audio latents are produced by independently pretrained VAEs, so cross-modal correspondence must either be learned from scratch by the generator or injected as external conditioning features. This motivates OmniVAE: we build cross-modal correspondence directly into the tokenizer by aligning the audio and video latent spaces, so that downstream joint generators operate on latents that already encode it.

%% file: chapters/method.tex
\section{OmniVAE}

\subsection{Overview}
OmniVAE retains separate encoders and decoders for each modality. The
video branch encodes
$x^v \in \mathbb{R}^{(1+T) \times H \times W \times 3}$ into a latent
$z^v \in \mathbb{R}^{C_v \times T_v \times (H/8) \times (W/8)}$ with
$4{\times}$ temporal and $8{\times}$ spatial downsampling, where
$T_v = 1 + T/4$ denotes the number of latent frames; the audio branch
encodes a 48\,kHz waveform $x^a \in \mathbb{R}^{1 \times L}$ into a 1D
latent $z^a \in \mathbb{R}^{C_a \times T_a}$, where $T_a = L / d_a$
for a temporal downsampling factor $d_a$. The two branches are
architecturally independent and do not interact at inference, so they
can be deployed jointly or used on their own. The video and audio
branches instantiate the Wan2.2 VAE backbone~\citep{wan22modelcard} and a
DAC-style continuous VAE~\citep{kumar2023dac}.

On top of these two VAEs, we introduce two training-only objectives.
During training we use paired samples $(x^v, x^a)$ extracted from the
same temporal window of a single source video, so that the two streams
are naturally aligned in both semantics and time. A \emph{segment-level
audio-video contrastive} objective aligns $z^v$ and $z^a$ at fine
temporal granularity in a shared embedding space, providing the
downstream generator with a latent representation in which cross-modal
correspondence is already established and reducing the burden of jointly
modeling both modalities. In parallel, a \emph{per-modality semantic
distillation} objective supervises each branch with features from a
frozen pretrained semantic encoder, making each branch's latent space
easier to model on its own. Neither objective is active at inference
time, so OmniVAE preserves the standard VAE interface and incurs no
additional runtime cost.

\subsection{Segment-Level Audio-Video Alignment}

Our goal is to make fine-grained audio-video correspondence an
intrinsic property of the VAE representations. Inspired by the
segment-level contrastive formulation of
Synchformer~\citep{iashin2024synchformer}, we map the heterogeneous
audio and video latents into temporally corresponding segment
representations and contrast matched segments against negatives at
multiple levels. The resulting contrastive objective encourages both
VAE encoders to preserve fine-grained audio-video correspondence in
their latent representations.

\noindent\textbf{Segment Representation.}
Given a paired audio-video clip, two modality-specific aggregation
paths map the heterogeneous latents to sequences of $S$ temporally
aligned features in a shared $d$-dimensional space. For video, the
latent $z_i^v \in \mathbb{R}^{C_v \times T_v \times H_v \times W_v}$
is first spatially aggregated into a frame sequence in
$\mathbb{R}^{(T_v-1) \times d_v}$ and then temporally pooled and
projected into $u_i^v \in \mathbb{R}^{S \times d}$. We exclude the first
latent frame because, in the causal video VAE, it covers a different
temporal extent from the remaining frames. For audio, the latent
$z_i^a \in \mathbb{R}^{C_a \times T_a}$ is transformed along the
temporal dimension and pooled directly into
$u_i^a \in \mathbb{R}^{S \times d}$. We implement the video spatial
aggregator with a lightweight Transformer and the audio temporal
aggregator with ConvNeXt-1D blocks. We write their segment features as
$u^v_{i,s},u^a_{i,s}\in\mathbb{R}^{d}$, where
$i$ indexes the clip and $s\in\{1,\ldots,S\}$ its temporal segment.

\noindent\textbf{Hierarchical Negative Construction.}
Learning fine-grained audio-video correspondence requires a negative
pool that combines hard temporal distinctions with broad semantic
diversity. Our preprocessing pipeline divides each long source video
into 8-second clips and further partitions each clip into $S=48$
segments of $1/6$ second. Synchformer draws negatives only from the
other segments within the anchor clip and from the segments of one
additional clip~\citep{iashin2024synchformer}. In contrast, we retain
the source-video identity of each clip and exploit the resulting
source-video--clip--segment hierarchy together with the distributed
batch to construct negatives at three levels. For an anchor segment
$(i,s)$, its paired segment in the other modality is the positive,
while the negative set $\mathcal{N}_{i,s}$ contains:
\textbf{(i) intra-clip negatives}, the other $S-1$ segments in the same
clip; \textbf{(ii) sibling-clip negatives}, segments from a temporally
disjoint clip of the same source video; and \textbf{(iii) cross-video
negatives}, segments sampled from unrelated source videos across the
distributed batch. In the default setting, we use all $S-1=47$
available intra-clip negatives and sample fixed numbers of 48
sibling-clip and 24 cross-video negatives, yielding an approximately
$2{:}2{:}1$ ratio across the three sources. The first two groups retain
similar content and context while differing in temporal alignment,
encouraging fine-grained temporal discrimination; the third expands
semantic diversity. We use the same construction in both contrastive
directions. The effects of clip duration and segment granularity are
studied in Sec.~\ref{sec:ablation_clip}.

\noindent\textbf{Bidirectional Contrastive Objective.}
Let $\mathcal{P}_{i,s}=\{(i,s)\}\cup\mathcal{N}_{i,s}$ be the candidate
set for anchor $(i,s)$ and define
$\mathrm{sim}_{i,s,j,t}=u^v_{i,s}\cdot u^a_{j,t}$. Because the features
are normalized, this dot product is cosine similarity. For a batch of
$B$ paired clips and a learnable temperature $\tau$, the
video-to-audio InfoNCE loss~\citep{oord2018cpc} is
\begin{equation}
\small
\mathcal{L}_{v \rightarrow a} = -\frac{1}{BS}\sum_{i=1}^{B}\sum_{s=1}^{S}
\log \frac{\exp(\mathrm{sim}_{i,s,i,s} / \tau)}{\sum_{(j,t) \in \mathcal{P}_{i,s}}\exp(\mathrm{sim}_{i,s,j,t} / \tau)}.
\end{equation}
$\mathcal{L}_{a \rightarrow v}$ is defined analogously by swapping the
two modalities, and the final alignment loss is
$\mathcal{L}_{\mathrm{avclip}}=\tfrac{1}{2}(\mathcal{L}_{v \rightarrow a}
+\mathcal{L}_{a \rightarrow v})$.

\subsection{Modality-Specific Semantic Distillation}

Semantic supervision from pretrained encoders can make VAE latents
easier for downstream generative models to learn~\citep{yu2024repa,
leng2025repae, singh2025irepa}. We therefore distill semantic features
into the audio and video branches independently, improving the
structure of each latent space without requiring cross-modal inputs.

\noindent\textbf{Teacher Selection.}
Our data include both images and videos on the visual side, and both
general audio and speech on the audio side. Qwen3-Omni~\citep{qwen3omni}
provides a visual encoder shared across images and videos and an audio
encoder that covers both general audio and speech. We freeze these two
encoders and use them as the video and audio teachers, respectively.

\noindent\textbf{Feature Distillation.}
For each modality, a lightweight projector maps the VAE latent to the
channel dimension and temporal resolution of its teacher features,
using convolution for channel projection and average pooling for
temporal alignment. For video, the first causal latent frame corresponds
only to the first input frame, so we align it with the teacher's image
features and align the remaining latent frames with its video features.
The audio branch uses a 1D convolutional projector. Following
iREPA~\citep{singh2025irepa}, we spatially normalize the visual teacher
features to emphasize local variation over the dominant global
component. At each aligned position $j$, we apply a sigmoid-cosine loss
between the projected latent $\tilde{z}_j$ and teacher feature $f_j$:
\begin{equation}
\mathcal{L}_{\mathrm{cos}} = -\frac{1}{|J|} \sum_{j \in J}
\log \sigma\!\left( \cos(\tilde{z}_j, f_j) \right),
\end{equation}
where $J$ indexes spatio-temporal positions for video and temporal
positions for audio. This yields the video and audio distillation losses
$\mathcal{L}_{\mathrm{vd}}$ and $\mathcal{L}_{\mathrm{ad}}$, respectively.

\subsection{Training Objective}

\noindent\textbf{Loss Formulation.}
We define separate objectives for the video and audio VAEs. Each
contains its modality-specific reconstruction and distillation losses,
together with the shared segment-level contrastive loss:
\begin{equation}
\begin{aligned}
\mathcal{L}_{v} &=
\lambda_{\mathrm{vr}}\mathcal{L}_{\mathrm{vr}}
+\lambda_{\mathrm{vd}}\mathcal{L}_{\mathrm{vd}}
+\lambda_{\mathrm{avclip}}\mathcal{L}_{\mathrm{avclip}}, \\
\mathcal{L}_{a} &=
\lambda_{\mathrm{ar}}\mathcal{L}_{\mathrm{ar}}
+\lambda_{\mathrm{ad}}\mathcal{L}_{\mathrm{ad}}
+\lambda_{\mathrm{avclip}}\mathcal{L}_{\mathrm{avclip}}.
\end{aligned},
\end{equation}
where $\mathcal{L}_{\mathrm{vr}}$ and $\mathcal{L}_{\mathrm{ar}}$ are
the video and audio reconstruction losses, and
$\mathcal{L}_{\mathrm{vd}}$ and $\mathcal{L}_{\mathrm{ad}}$ are their
distillation losses. The reconstruction terms comprise L1,
LPIPS~\citep{zhang2018lpips}, and KL for video, and waveform,
multi-resolution mel, and KL losses for audio. The shared
$\mathcal{L}_{\mathrm{avclip}}$ appears in both objectives because it
supervises both encoders, but is evaluated only once during joint
optimization. Since these loss groups differ substantially in scale, we
set the coefficients $\lambda$ using loss-magnitude weighting to keep
their contributions comparable during training.

\noindent\textbf{Training Schedule.}
We use a three-stage schedule to accommodate the different convergence
rates and memory requirements of reconstruction and semantic learning.
\textbf{Stage~1} independently pretrains the two VAEs using only
reconstruction objectives, allowing them to acquire basic reconstruction
capability through substantially more training steps than required by
the semantic objectives. Both branches use adversarial supervision, and
the audio branch additionally uses discriminator feature matching.
\textbf{Stage~2} introduces contrastive alignment and semantic
distillation. Contrastive learning requires long clips and a large
negative pool, making this stage particularly memory intensive. We
therefore remove the discriminators and feature matching, and jointly
train both VAEs with the contrastive head and distillation projectors.
Removing the
audio discriminator, however, can introduce mild electronic artifacts
in reconstructed audio. In \textbf{Stage~3}, we consequently freeze the
audio encoder and fine-tune only its decoder with the Stage~1 audio
losses, restoring adversarial and feature-matching supervision. This
refinement improves perceptual audio quality without altering the
aligned latent space; the video branch remains unchanged.

\subsection{Downstream Generation}

To verify that OmniVAE's latent space benefits downstream generation, we
freeze the trained OmniVAE as a tokenizer and build a joint
text-to-audio-video (T2AV) model on top of it, trained with flow
matching in the video and audio latent spaces. Following common
practice, we proceed in two stages: we first pretrain strong
single-modality priors—a visual branch (T2I/T2V) and an audio branch
(T2A)—and then couple them through lightweight cross-attention bridges
and train jointly with the per-modality flow-matching losses. The model
supports both joint and single-modality decoding at inference.

%% file: chapters/exp.tex
\section{Experiments}
\label{sec:exp}

\subsection{Experimental Setup}
\label{sec:exp_setup}

\noindent\textbf{Datasets.}
Existing large-scale audio-visual datasets are dominated by human
speech. To enrich the model's knowledge of the associations between
non-speech sound events and their visual sources, we devise a
top-down data collection pipeline. We first use an LLM to construct
an audio-visual concept tree, which is subsequently verified by human
annotators. For each leaf concept, the LLM then generates multiple
search queries to retrieve videos whose visual and acoustic content
matches the concept. This pipeline substantially broadens the coverage
and diversity of sound events and audio-visual concepts in our corpus.
Web-sourced videos, however, are noisy and often exhibit weak
visual-audio correlation---e.g., background music or first-person
voice-overs decoupled from the on-screen content. To obtain data with
fine-grained audio-visual correspondence, we further apply a
filtering pipeline that combines audio-visual semantic alignment
scoring~\citep{girdhar2023imagebind}, DeSync-based synchronization
filtering~\citep{iashin2024synchformer}, and audio event
detection~\citep{chen2023beats}.
Combined with AudioSet~\citep{gemmeke2017audioset}, the resulting
corpus contains about $23$M audio-visual clips, evenly split between
speech and non-speech samples. Joint generation training further
requires three types of captions---visual, audio, and unified
audio-visual---for which we use Qwen3.5-VL~\citep{qwen35} for visual
captions and two in-house fine-tuned variants of
Qwen3-Omni~\citep{qwen3omni} for the audio and unified audio-visual
captions, respectively.

\noindent\begin{tabularx}{\textwidth}{@{}p{0.48\textwidth}@{\hspace{0.04\textwidth}}p{0.46\textwidth}@{}}
\begin{minipage}[t]{\linewidth}
\vspace{0pt}\justifying
\noindent\textbf{Model.} For semantic distillation, we use features from layer $27$
of the Qwen3-Omni~\citep{qwen3omni} visual encoder and layer $18$ of
its audio encoder as supervision. In terms of model size, the two VAEs
comprise approximately $1.08$B parameters and are the only components
retained at inference. The contrastive head and distillation projectors
add approximately $49$M training-only parameters and are discarded after
training, introducing no inference overhead. Table~\ref{tab:params}
provides the detailed parameter breakdown.
\end{minipage}
&
\begin{minipage}[t]{\linewidth}
\vspace{0pt}\centering\small
\captionsetup{skip=3pt}
\setlength{\tabcolsep}{12pt}
\renewcommand{\arraystretch}{0.95}
\begin{tabular*}{\linewidth}{@{\extracolsep{\fill}}lr@{}}
\toprule
\textbf{Module} & \textbf{\#Params} \\
\midrule
\multicolumn{2}{l}{\textit{Retained at inference}} \\
\quad Video VAE (enc. + dec.) & 704.7\,M \\
\quad Audio VAE (enc. + dec.) & 371.6\,M \\
\quad \textit{Subtotal} & \textit{1{,}076.3\,M} \\
\midrule
\multicolumn{2}{l}{\textit{Training-only}} \\
\quad Contrastive head & 20.5\,M \\
\quad Distill. projectors & 29.0\,M \\
\quad \textit{Subtotal} & \textit{49.5\,M} \\
\midrule
\textbf{Total (training)} & \textbf{1{,}125.8\,M} \\
\bottomrule
\end{tabular*}
\captionof{table}{Parameter counts. Training-only modules are discarded at inference.}
\label{tab:params}
\end{minipage}
\end{tabularx}
\par\vspace{0.4em}

\noindent\textbf{Training and Baselines.}
We organize the experiments as a progressive ablation over training
objectives that isolates the contribution of reconstruction, semantic
distillation, and audio-visual contrastive learning:
\textbf{(i) Recon}---only the per-modality reconstruction losses,
with the video and audio VAEs trained independently;
\textbf{(ii) Recon+Distill}---adds modality-specific semantic distillation,
still training the two VAEs independently;
\textbf{(iii) Recon+AVCLIP}---instead adds the segment-level
audio-visual contrastive loss and trains the two VAEs jointly; and
\textbf{(iv) OmniVAE} (full model)---trains the two VAEs jointly with
reconstruction, distillation, and contrastive objectives enabled
simultaneously. The video and audio branches of OmniVAE are
initialized from the Recon+Distill checkpoints, while the contrastive
head is trained from scratch.

The per-modality variants (i, ii) are trained for $250$k steps on
$121$-frame clips, and the joint variants (iii, iv) for $82$k steps
on $193$-frame clips with a global batch size of $256$. Joint
training follows a two-stage schedule that first freezes the video
VAE to prioritize audio reconstruction and cross-modal alignment,
and then unfreezes it to jointly optimize all objectives. Since joint
training omits the audio discriminators for memory efficiency, we
briefly fine-tune the audio decoder of the AVCLIP-based variants with
adversarial supervision, removing mild electronic artifacts with
negligible changes in objective metrics.
The relative scales of the loss groups follow the
contrastive-anchored loss-magnitude weighting discussed in
\S\ref{sec:ablation_loss}. All variants use the same video and audio
VAE backbones, training data, and optimization settings.

\subsection{Reconstruction Evaluation}
\label{sec:recon_eval}

We first compare the reconstruction quality of different training
configurations. The video VAE is evaluated on
UCF-101~\citep{soomro2012ucf101} and
Panda-70M~\citep{chen2024panda70m} at $24$\,fps, $121$ frames, and
$256 \times 256$. The audio VAE is evaluated on
LibriSpeech~\citep{panayotov2015librispeech} test-clean for speech,
AudioSet~\citep{gemmeke2017audioset} for general audio, and
MUSDB18~\citep{rafii2017musdb18} for music.

Table~\ref{tab:vae_recon} reports the results. On the video side,
all training variants remain close to the reconstruction-only baseline
across the reconstruction metrics, indicating that the additional
objectives largely preserve video reconstruction quality. Our Recon
model is initialized from the official Wan2.2 VAE checkpoint and then
fine-tuned using only reconstruction objectives. The official Wan VAEs are
trained across diverse video settings and are therefore included for
reference only. On the audio side, our models build upon the open-source
HunyuanVideo-Foley VAE. Semantic distillation largely preserves
reconstruction quality, while contrastive alignment leads to only a
slight degradation. Even so, OmniVAE remains competitive with or
outperforms the external audio VAEs on most reported metrics. Overall, OmniVAE retains strong
reconstruction quality across video, speech, general audio, and music.


\begin{table*}[t]
\centering
\small
\setlength{\tabcolsep}{4pt}
\begin{tabular*}{\dimexpr\textwidth-1.5pt\relax}{@{\extracolsep{\fill}}lccccc@{}}
\toprule
\multicolumn{6}{@{}l}{\textbf{(a) Video Reconstruction} \quad \textit{UCF-101 / Panda-70M}} \\
\addlinespace[2pt]
\midrule
\hspace{0.6em}Configuration & PSNR $\uparrow$ & SSIM $\uparrow$ & LPIPS $\downarrow$ & L1 $\downarrow$ & rFVD $\downarrow$ \\
\midrule
\multicolumn{6}{@{}l}{\textit{External VAEs}} \\
\hspace{0.6em}Wan2.1 VAE~\citep{wan2025} & 34.50 / 32.62 & 0.9510 / 0.9448 & 0.0201 / 0.0169 & 0.0117 / 0.0130 & 2.46 / 1.49 \\
\hspace{0.6em}Wan2.2 VAE~\citep{wan22modelcard} & 34.87 / 33.21 & 0.9584 / 0.9541 & 0.0177 / 0.0137 & 0.0110 / 0.0120 & 3.72 / 1.38 \\
\midrule
\multicolumn{6}{@{}l}{\textit{Ours}} \\
\hspace{0.6em}Recon                    & 36.93 / 36.56 & 0.9656 / 0.9697 & 0.0102 / 0.0067 & 0.0090 / 0.0086 & 2.40 / 1.25 \\
\hspace{0.6em}Recon + Distill          & 36.70 / 36.22 & 0.9649 / 0.9688 & 0.0107 / 0.0071 & 0.0092 / 0.0089 & 2.53 / 1.26 \\
\hspace{0.6em}Recon + AVCLIP           & 36.22 / 35.13 & 0.9616 / 0.9631 & 0.0115 / 0.0079 & 0.0097 / 0.0098 & 2.66 / 1.35 \\
\hspace{0.6em}OmniVAE                  & 36.27 / 35.24 & 0.9616 / 0.9632 & 0.0113 / 0.0077 & 0.0097 / 0.0098 & 2.81 / 1.31 \\
\bottomrule
\end{tabular*}
\vspace{0.8em}
\renewcommand{\arraystretch}{0.95}
\begin{tabular*}{\dimexpr\textwidth-1.5pt\relax}{@{}l@{\extracolsep{\fill}\hspace{-0.4em}}c@{\extracolsep{0pt}\hspace{1.5em}}c@{\hspace{1.5em}}c@{\hspace{1.5em}}c@{\extracolsep{\fill}\hspace{-0.4em}}c@{\extracolsep{0pt}\hspace{1.4em}}c@{\hspace{1.4em}}c@{}}
\multicolumn{8}{@{}l}{\textbf{(b) Audio Reconstruction}} \\
\addlinespace[2pt]
\toprule
& \multicolumn{4}{c}{Speech} & \multicolumn{3}{c}{Audio / Music} \\
\cmidrule(lr){2-5}\cmidrule(l){6-8}
\hspace{0.6em}Model & SIM $\uparrow$ & STOI $\uparrow$ & P-NB $\uparrow$ & P-WB $\uparrow$ & Mel $\downarrow$ & STFT $\downarrow$ & ViSQOL $\uparrow$ \\
\midrule
\multicolumn{8}{@{}l}{\textit{External VAEs}} \\
\hspace{0.6em}HunyuanVideo-Foley VAE~\citep{shan2025hunyuanvideofoley} & 0.9899 & 0.9979 & 4.4780 & 4.5337 & 0.52 / 0.60 & 1.82 / 1.84 & 4.47 / 4.45 \\
\hspace{0.6em}MMAudio VAE~\citep{cheng2025mmaudio} & 0.9122 & 0.9502 & 3.4931 & 2.9797 & 1.23 / 1.36 & 2.14 / 2.44 & 4.42 / 4.42 \\
\hspace{0.6em}Stable Audio 3 SAME-S~\citep{evans2026stableaudio3} & 0.7960 & 0.9139 & 2.8073 & 2.2991 & 1.46 / 1.37 & 3.19 / 3.67 & 3.68 / 3.99 \\
\hspace{0.6em}Stable Audio 3 SAME-L~\citep{evans2026stableaudio3} & 0.8755 & 0.9539 & 3.4615 & 3.0739 & 1.37 / 1.22 & 3.30 / 3.55 & 3.76 / 4.17 \\
\midrule
\multicolumn{8}{@{}l}{\textit{Ours}} \\
\hspace{0.6em}Recon & 0.9893 & 0.9981 & 4.4916 & 4.5557 & 0.44 / 0.50 & 1.75 / 1.72 & 4.55 / 4.60 \\
\hspace{0.6em}Recon + Distill & 0.9893 & 0.9970 & 4.4453 & 4.5101 & 0.56 / 0.64 & 1.96 / 1.95 & 4.44 / 4.42 \\
\hspace{0.6em}Recon + AVCLIP & 0.9666 & 0.9946 & 4.4346 & 4.3638 & 0.76 / 0.75 & 1.99 / 2.03 & 4.03 / 4.18 \\
\hspace{0.6em}OmniVAE & 0.9737 & 0.9948 & 4.4400 & 4.4085 & 0.74 / 0.73 & 2.02 / 2.08 & 4.08 / 4.31 \\
\bottomrule
\end{tabular*}
\caption{Video and audio reconstruction quality. Video values follow
UCF-101 / Panda-70M, while Audio / Music values follow AudioSet / MUSDB18.
All our video variants are
initialized from Wan2.2, and our audio variants from the
HunyuanVideo-Foley VAE. External models serve as reference points rather
than training-matched baselines.}
\label{tab:vae_recon}
\end{table*}

\subsection{Audio-Video Sync Probing}
\label{sec:sync_probing}

Evaluating fine-grained audio-video alignment through downstream
joint generation is expensive, as each comparison requires training
a full generative model. We therefore adopt audio-video sync probing
as an efficient and sensitive proxy. The probe predicts the temporal
offset between audio and video, requiring it to associate sound events
with their visual sources at fine temporal granularity.
\par\medskip
\noindent\begin{tabularx}{\textwidth}{@{}p{0.46\textwidth}@{\hspace{0.04\textwidth}}p{0.48\textwidth}@{}}
\begin{minipage}[t]{\linewidth}
\vspace{0pt}\justifying
We evaluate on VGGSound-Sparse (VGS-Sp), which contains sparse,
temporally localizable events~\citep{chen2020vggsound,iashin2022sparse,
iashin2024synchformer}. Following Synchformer, we shift audio by
$N\times0.2$ seconds and train a lightweight head to classify 21 offsets
from $-2$ to $+2$ seconds. We report A@1, A@5, and A@1$_{\mathrm{tol}}$
with a $\pm1$-class tolerance. Under \textbf{frozen}, only the classifier
is trained; \textbf{+ft} also tunes the contrastive aggregators.
Table~\ref{tab:sync_probing} shows that contrastive learning provides
the main gain in temporal alignment, with OmniVAE performing best under
both protocols.
\end{minipage}
&
\begin{minipage}[t]{\linewidth}
\vspace{0pt}\centering
\small
\captionsetup{skip=1pt,font=small}
\setlength{\tabcolsep}{7.5pt}
\renewcommand{\arraystretch}{0.76}
\begin{tabular}{@{}lccc@{}}
\toprule
\textbf{Method} & \textbf{A@1} & \textbf{A@5} &
\textbf{A@1$_{\text{tol}}$} \\
\midrule
\multicolumn{4}{@{}l}{\textit{Frozen}} \\
Recon                  & 6.4  & 31.5 & 16.2 \\
Recon + Distill        & 7.1  & 33.0 & 17.5 \\
Recon + AVCLIP         & 18.1 & 47.8 & 34.5 \\
\textbf{OmniVAE (ours)} & \textbf{20.2} & \textbf{50.1} & \textbf{37.2} \\
\midrule
\multicolumn{4}{@{}l}{\textit{+ft}} \\
Recon $+$ ft                 & 7.0  & 32.4 & 17.1 \\
Recon + Distill $+$ ft       & 7.8  & 33.9 & 18.4 \\
Recon + AVCLIP $+$ ft        & 21.8 & 52.4 & 39.6 \\
\textbf{OmniVAE $+$ ft}       & \textbf{22.3} & \textbf{53.2} & \textbf{40.4} \\
\bottomrule
\end{tabular}
\captionof{table}{Sync probing on VGS-Sp (\%); A@1$_{\mathrm{tol}}$
uses $\pm1$ tolerance.}
\label{tab:sync_probing}
\end{minipage}
\end{tabularx}
\par\vspace{0.4em}

\subsection{Text-to-Audio-Video Generation}
\label{sec:t2av}

\begin{figure}[t]
    \centering
    \includegraphics[width=0.99\linewidth]{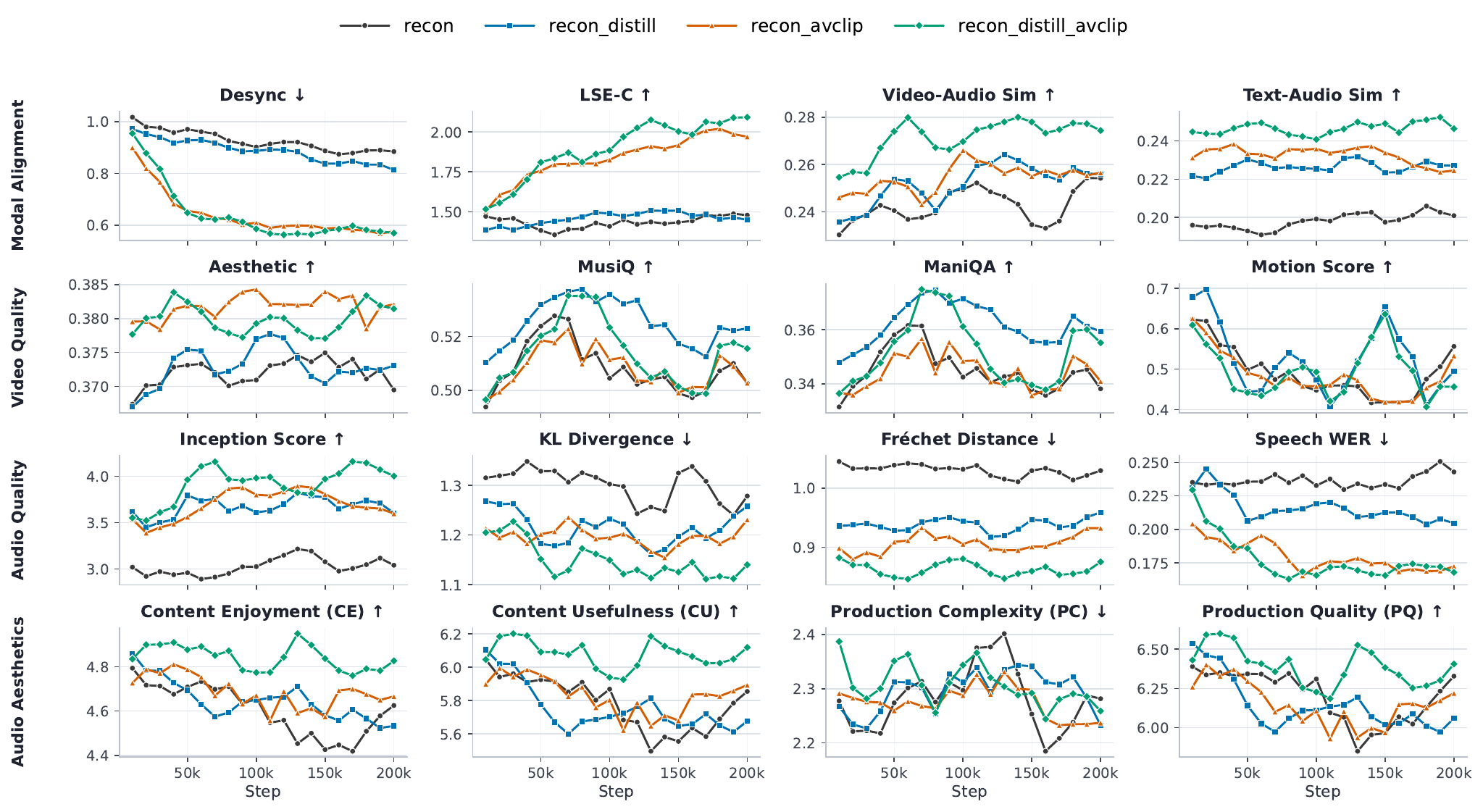}
    \caption{Evaluation metrics of T2AV models trained on the four
    VAE configurations, tracked across training steps. Rows, top to
    bottom: cross-modal alignment (Desync, LSE-C, Video-Audio Sim,
    Text-Audio Sim), video quality (Aesthetic, MusiQ, ManiQA, Motion
    Score), AudioBox aesthetics~\citep{tjandra2025audiobox} (Content
    Enjoyment, Content Usefulness, Production Complexity, Production
    Quality), and audio quality (IS, KL Divergence, Fr\'echet
    Distance, Speech WER). The AVCLIP-equipped configurations
    consistently lead on all cross-modal alignment metrics, both
    semantic objectives improve audio quality, and the full OmniVAE
    (\textit{recon\_distill\_avclip}) achieves the best overall
    results.}
    \label{fig:t2av_curves}
\end{figure}

\noindent\textbf{Benchmark.}
We evaluate T2AV generation on Verse-Bench~\citep{ye2025universe}
using the unified T2AV prompts introduced in MOVA~\citep{mova}.
Verse-Bench comprises three subsets: Set1-I ($205$
samples sourced from still images), Set2-V ($295$ samples sourced
from web videos), and Set3-TED ($100$ TED-talk samples used for
lip-sync evaluation). Because our T2AV models generate at
$256 \times 256$ resolution, faces are often too small for
SyncNet~\citep{chung2017syncnet} to detect, leaving only a small
number of samples for which a lip-sync score can be computed. We
therefore use an LLM to rewrite each Set3 visual prompt into a
medium close-up or close-up shot while preserving its original
intent and speech content, so that faces occupy a larger fraction of
the frame. All baselines are evaluated on the rewritten prompts for
fair comparison. Set3 results under this protocol are not directly
comparable to those reported using the original Verse-Bench prompts.

\noindent\textbf{Results.}
Figure~\ref{fig:t2av_curves} compares the four VAE configurations
throughout T2AV training. For a checkpoint-independent quantitative
comparison, Table~\ref{tab:t2av_final} reports the mean over the final
three checkpoints (180k, 190k, and 200k) under CFG${}=5$, using the same
checkpoints for every configuration. On \textbf{cross-modal alignment}, Desync
measures fine-grained temporal synchronization and LSE-C measures
lip synchronization, while Video-Audio Sim and Text-Audio Sim
evaluate overall semantic consistency. The AVCLIP-based variants
consistently improve both groups of metrics, with OmniVAE achieving
the strongest overall alignment.
On \textbf{audio quality}, the AudioBox scores jointly assess content
and production quality, IS, KL divergence, and Fr\'echet distance
evaluate non-speech audio generation, and WER measures speech
intelligibility. Both semantic distillation and AVCLIP improve these
metrics individually, while their combination achieves the strongest
overall audio performance.
On \textbf{video quality}, AVCLIP improves visual aesthetics, while
semantic distillation yields gains on MusiQ and ManiQA; motion quality
remains broadly comparable across configurations.
Together, these results show that AVCLIP improves both cross-modal
alignment and audio generation quality, while semantic distillation
further strengthens modality-specific generation quality. Combining
the two objectives yields complementary gains and the strongest
overall performance.

\begin{table*}[t]
\centering
\small
\setlength{\tabcolsep}{5pt}
\renewcommand{\arraystretch}{1.05}
\begin{tabular*}{\textwidth}{@{\extracolsep{\fill}}lcccc@{}}
\toprule
\multicolumn{5}{@{}l}{\textbf{Cross-Modal Alignment}} \\
\addlinespace[2pt]
Configuration & DeSync $\downarrow$ & LSE-C $\uparrow$ & Video-Audio Sim $\uparrow$ & Text-Audio Sim $\uparrow$ \\
\midrule
Recon                  & 0.884 & 1.479 & 0.254 & 0.201 \\
Recon + Distill        & 0.814 & 1.450 & 0.256 & 0.227 \\
Recon + AVCLIP         & 0.576 & 1.970 & 0.257 & 0.225 \\
OmniVAE                & \textbf{0.570} & \textbf{2.093} & \textbf{0.274} & \textbf{0.246} \\
\midrule
\multicolumn{5}{@{}l}{\textbf{Video Quality}} \\
\addlinespace[2pt]
Configuration & Aesthetic $\uparrow$ & MusiQ $\uparrow$ & ManiQA $\uparrow$ & Motion Score $\uparrow$ \\
\midrule
Recon                  & 0.369 & 0.503 & 0.338 & \textbf{0.556} \\
Recon + Distill        & 0.373 & \textbf{0.523} & \textbf{0.359} & 0.495 \\
Recon + AVCLIP         & \textbf{0.382} & 0.503 & 0.341 & 0.533 \\
OmniVAE                & 0.381 & 0.515 & 0.355 & 0.456 \\
\midrule
\multicolumn{5}{@{}l}{\textbf{Audio Quality}} \\
\addlinespace[2pt]
Configuration & IS $\uparrow$ & KL $\downarrow$ & FD $\downarrow$ & WER $\downarrow$ \\
\midrule
Recon                  & 3.041 & 1.279 & 1.029 & 0.243 \\
Recon + Distill        & 3.604 & 1.258 & 0.959 & 0.205 \\
Recon + AVCLIP         & 3.598 & 1.231 & 0.932 & 0.172 \\
OmniVAE                & \textbf{4.001} & \textbf{1.140} & \textbf{0.875} & \textbf{0.168} \\
\midrule
\multicolumn{5}{@{}l}{\textbf{AudioBox Aesthetics}} \\
\addlinespace[2pt]
Configuration & CE $\uparrow$ & CU $\uparrow$ & PC $\downarrow$ & PQ $\uparrow$ \\
\midrule
Recon                  & 4.625 & 5.855 & 2.282 & 6.328 \\
Recon + Distill        & 4.533 & 5.677 & \textbf{2.232} & 6.060 \\
Recon + AVCLIP         & 4.666 & 5.894 & 2.237 & 6.219 \\
OmniVAE                & \textbf{4.826} & \textbf{6.119} & 2.259 & \textbf{6.406} \\
\bottomrule
\end{tabular*}
\caption{T2AV generation results averaged over the final three checkpoints
(180k, 190k, and 200k) with CFG${}=5$. All metrics are aggregated over the
supported Verse-Bench subsets; LSE-C is evaluated on the face-detectable
Set3 samples. CE, CU, PC, and PQ denote Content Enjoyment, Content
Usefulness, Production Complexity, and Production Quality, respectively.
Bold indicates the best result for each metric.}
\label{tab:t2av_final}
\end{table*}

\subsection{Ablation Study}
\label{sec:ablation}

We ablate two key designs: the temporal granularity of contrastive
learning (\S\ref{sec:ablation_clip}) and the loss-balancing strategy
(\S\ref{sec:ablation_loss}). We evaluate alignment using sync probing
for the former and segment-retrieval accuracy for the latter, while
also reporting reconstruction quality.

\subsubsection{Contrastive Pool Size and Temporal Granularity}
\label{sec:ablation_clip}
\noindent\begin{minipage}[t]{0.52\textwidth}
\vspace{0pt}
\justifying
The contrastive pool is determined by the video frame rate and clip
duration, while the segment length controls temporal granularity. As
shown in Table~\ref{tab:ablation_clip}, using a higher frame rate
together with finer temporal segments, or increasing the clip duration,
substantially improves sync probing accuracy, demonstrating the importance
of a sufficiently large negative pool.
By contrast, performance is relatively stable across segment lengths.
We therefore use $24$\,fps, $8$-second clips, and $0.17$-second segments
in the main experiments, which provide a large negative pool at the
finest evaluated temporal granularity.
\end{minipage}\hfill
\begin{minipage}[t]{0.44\textwidth}
\vspace{0pt}
\centering
\small
\captionsetup{skip=3pt}
\setlength{\tabcolsep}{10.5pt}
\renewcommand{\arraystretch}{1.08}
\begin{tabular}{@{}cccc@{}}
\toprule
\textbf{FPS} & \textbf{Clip} & \textbf{Segment} & \textbf{A@1 (\%)} \\
\midrule
 8 & 8\,s   & 0.50\,s & 19.0 \\
24 & ---    & 0.17\,s & 7.0 \\
24 & 2\,s   & 0.17\,s & 6.0 \\
24 & 8\,s   & 0.17\,s & \textbf{30.0} \\
24 & 8\,s   & 0.34\,s & 27.0 \\
24 & 8\,s   & 0.50\,s & 29.0 \\
\bottomrule
\end{tabular}
\captionof{table}{Ablation of contrastive pool size and temporal granularity.
A@1 is measured using the VGS-Sp temporal-offset probe in
Table~\ref{tab:sync_probing}. The dash denotes the reconstruction-only
baseline without contrastive training.}
\label{tab:ablation_clip}
\end{minipage}
\par\vspace{0.8em}
\subsubsection{Balancing Reconstruction and Cross-Modal Alignment}
\label{sec:ablation_loss}

Joint training combines video reconstruction, audio reconstruction,
and contrastive alignment losses with substantially different scales.
Without appropriate balancing, the dense reconstruction objectives
can overwhelm the comparatively sparse alignment signal. We therefore
compare three strategies for balancing these loss groups:
loss-magnitude weighting, gradient-based weighting using the last
encoder layer as a proxy, and uncertainty-based
weighting~\citep{kendall2018multi}.

Because all strategies use the same contrastive setup, we evaluate
alignment through segment retrieval. \textbf{Intra A@48} retrieves
the paired segment within the same video, whereas \textbf{Overall
A@64} additionally includes negatives sampled from other videos in
the batch.
As shown in Table~\ref{tab:ablation_loss}, all three strategies
achieve comparable reconstruction quality, but loss-magnitude
weighting performs substantially better on both retrieval metrics.
We therefore adopt it for joint training.

\begin{table}[H]
\centering
\small
\setlength{\tabcolsep}{4pt}
\begin{tabular*}{\textwidth}{@{\extracolsep{\fill}}lcccccc@{}}
\toprule
\multirow{2}{*}{\textbf{Balancing strategy}} & \multicolumn{2}{c}{\textbf{Segment retrieval}} & \multicolumn{3}{c}{\textbf{Audio recon.}} & \textbf{Video recon.} \\
\cmidrule(lr){2-3} \cmidrule(lr){4-6} \cmidrule(lr){7-7}
& \textbf{Intra A@48} & \textbf{Overall A@64} & \textbf{STOI} & \textbf{PESQ-WB} & \textbf{PESQ-NB} & \textbf{PSNR} \\
\midrule
Loss-magnitude (ours)           & \textbf{0.51} & \textbf{0.24} & 0.994 & 4.41 & 4.38 & 34.1 \\
GradNorm (last-layer proxy)     & 0.44 & 0.17 & \textbf{0.995} & 4.42 & 4.40 & \textbf{35.4} \\
Uncertainty-based adaptive      & 0.29 & 0.13 & 0.994 & \textbf{4.45} & \textbf{4.41} & 33.4 \\
\bottomrule
\end{tabular*}
\caption{Ablation of loss-balancing strategies. Segment retrieval
measures fine-grained alignment, while STOI, PESQ, and PSNR measure
reconstruction quality.}
\label{tab:ablation_loss}
\end{table}

%% file: chapters/conclusions.tex
\section{Conclusion}

We presented OmniVAE, a jointly trained audio-video VAE that embeds modality-specific semantics and fine-grained cross-modal correspondence into the latent representations. OmniVAE augments standard reconstruction training with segment-level audio-video contrastive learning and modality-specific semantic distillation. Our experiments show that the two objectives provide complementary benefits. Contrastive learning substantially improves fine-grained audio-video alignment, while semantic distillation strengthens the learnability of the modality-specific representations. These enhanced representations enable downstream text-to-audio-video generation models to achieve better generation quality and more accurate cross-modal synchronization without materially compromising reconstruction quality. Taken together, these findings highlight the importance of incorporating both modality-specific semantic structure and cross-modal correspondence directly into learned representations, providing a stronger foundation for unified multimodal generation. Future work should explore more effective methods for modality-specific and cross-modal semantic learning and validate their benefits at larger scales.